\documentclass[aps,pra,showpacs,twocolumn]{revtex4-1}
\usepackage{amssymb}
\usepackage{amsmath}
\usepackage{graphicx}
\usepackage{epsfig}
\usepackage{subfigure}
\usepackage{amsfonts}
\usepackage{color}

\begin{document}

\title{Fixed lines in a non-Hermitian Kitaev chain with
spatially balanced pairing processes}
\author{Y. B. Shi}
\author{Z. Song}
\email{songtc@nankai.edu.cn}

\begin{abstract}
Exact solutions for non-Hermitian quantum many-body systems are rare but may
provide valuable insights into the interplay between Hermitian and
non-Hermitian components. We report our investigation of a non-Hermitian
variant of a $p$-wave Kitaev chain by introducing staggered imbalanced pair
creation and annihilation terms. We find that there exist fixed lines in
the phase diagram, at which the ground state remains unchanged in the
presence of non-Hermitian term under the periodic boundary condition for a
finite system.\ This allows the constancy of the topological index in the
process of varying the balance strength at arbitrary rate, exhibiting the
robustness of the topology for non-Hermitian Kitaev chain under
time-dependent perturbations. The underlying mechanism is investigated
through the equivalent quantum spin system obtained by the Jordan-Wigner
transformation for infinite chain. In addition, the exact solution shows
that a resonant non-Hermitian impurity can induce a pair of zero modes in
the corresponding Majorana lattice, which asymptotically approach the edge
modes in the thermodynamic limit, manifesting the bulk-boundary
correspondence. Numerical simulation is performed for the quench dynamics
for the systems with slight deviation from the fixed lines to show the
stability region in time. This work reveals the interplay between the pair
creation and annihilation pairing processes.
\end{abstract}

\maketitle
\affiliation{School of Physics, Nankai University, Tianjin 300071, China }

\section{Introduction}

The exact solution of a model Hamiltonian plays an important role in physics
and sometimes may open the door to the exploration of new frontiers in
physics. One recent example is the discovery of the solution for
non-Hermitian harmonic system, which is a starting point of $PT$-symmetric
quantum mechanics \cite{Bender,Bender1,Bender2,Bender3}. In traditional
quantum mechanics, the fundamental postulate of the Hermiticity of the
Hamiltonian ensures\ the reality of the spectrum and the unitary dynamics
for a closed quantum system \cite{DAMQ}. In general, any Hermitian
Hamiltonian can also be decomposed into two non-Hermitian sub-Hamiltonians
that are Hermitian conjugates of each other. Intuitively, the reality of the
spectrum is rooted in the balance of the actions of the two non-Hermitian
sub-Hamiltonians. In this sense, the Hermiticity is not necessary for the
balance, since the balance can be established across a distance by a
pseudo-Hermitian Hamiltonian \cite{Ali1,Ali2,Ali3,Ali4,Ali5,DMTN}, for
instance, a simple gain-loss-balanced system in Refs. \cite{Jin1,Jin2,Jin3}.
However, exact solutions for non-Hermitian quantum many-body systems are
rare but provide valuable insights into the physics of quantum
matter. In this work, we present a much more compelling example to
demonstrate the balance in a non-Hermitian quantum many-body system and
reveal the interplay between Hermitian and non-Hermitian components.

We study a non-Hermitian variant of a $p$-wave Kitaev chain \cite{Kitaev} by
introducing staggered imbalanced pair creation and annihilation terms. The
staggered arrangement provides the balance between two neighboring dimers. In most works, the balanced terms usually arise from
single-particle processes, such as hopping or on-site potential terms. In
previous work \cite{SYBPRB2}, it has been shown that a non-Hermitian term
for the pairing process can alter the phase diagram in the ground state. In
this work, the non-Hermitian components are extended. Based on
the exact solutions, we find that there exists a family of lines, at which
the ground state of the Hamiltonian remains unchanged in the presence of a
non-Hermitian term under the periodic boundary condition for a finite
system.\ The ground state is only determined by the intercept of a given
line, referred to as fixed line.

It is well known that the topological superconducting phase in the original
Kitaev model has been demonstrated by the winding number of the Majorana
lattice and unpaired Majorana modes exponentially localized at the ends of
open Kitaev chains \cite{Sarma,Stern,Alicea}. For the present model, our
result is the constancy of the winding number in the process of varying the
balance strength at an arbitrary rate, exhibiting the robustness of the
topology for a non-Hermitian Kitaev chain under time-dependent
perturbations. In addition, the exact solution shows that a resonant
non-Hermitian impurity can induce a pair of zero modes in the corresponding
Majorana lattice, which asymptotically approach the edge modes in the
thermodynamic limit, manifesting the bulk-boundary correspondence. The
underlying mechanism is investigated through the equivalent quantum spin
system \cite{SachdevBook} obtained by the Jordan-Wigner transformation \cite%
{Pfeuty} for finite chains. Starting from the explicit ground states of the
open spin chain, we elaborate that the non-Hermitian terms have no energy
contribution in the bulk, and show how the string-like boundary term
hybridizes two ferromagnetic states. We also investigate the stability
region in time for systems with a slight deviation from the fixed lines by
performing numerical simulation of quench dynamics. This work reveals the
interplay between the pair creation and annihilation pairing processes at
different locations.

This paper is organized as follows. In Section \ref{Model and fixed line},
we describe the model Hamiltonian and present the exact solution. In Section %
\ref{Majorana zero modes}, we study the corresponding Majorana lattice to
show the exact zero modes in the presence of an engineered impurity. In
Section \ref{Dissipationless ferromagnetic order}, we investigate our model
in the spin representation. In Section \ref{Stability of quench dynamics},
we present the features of the dynamic behavior when the\ quench
Hamiltonian\ deviates from the fixed lines. Finally, we give a summary and
discussion in Section \ref{summary}. Some details of derivations are placed
in the Appendix.

\begin{figure*}[t]
\centering\includegraphics[width=0.94\textwidth]{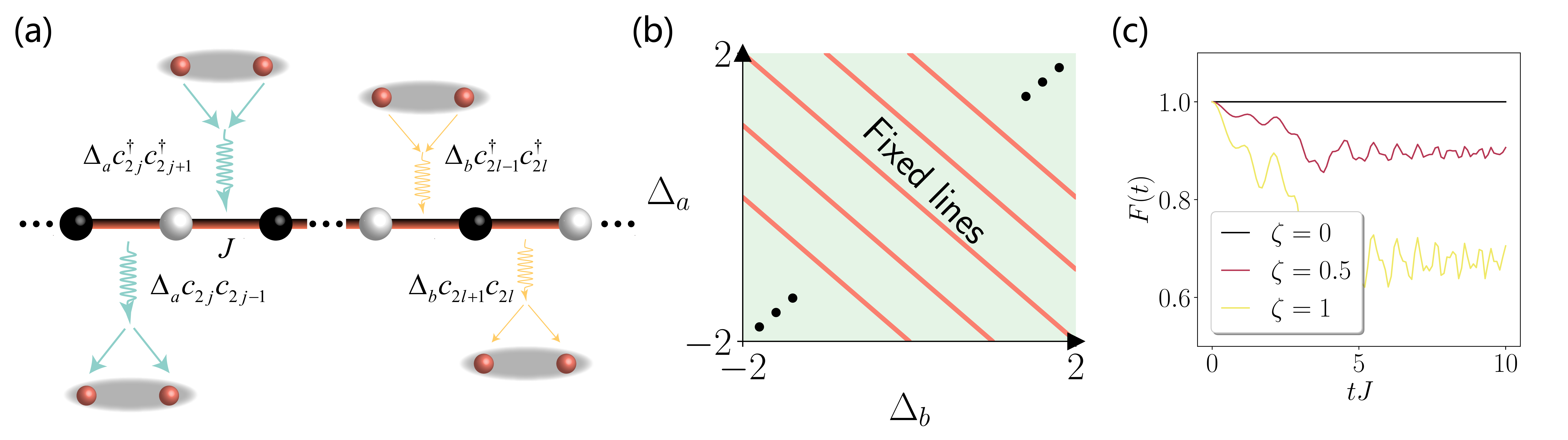}
\caption{(a) Schematic of a 1D Kitaev model for spinless fermion with
non-Hermitian imbalanced pair terms. It consists of two sublattices A and B
in black and white, respectively. Here, $J$ is the hopping strength between
two adjacent sites. $\Delta _{a}$ is the strength of the p-wave pair
creation (annihilation) on the odd (even) dimer while $\Delta _{b}$ is the
strength of the p-wave pair annihilation (creation) on the even (odd) dimer. 
(b) Schematic illustration of fixed lines. The fixed lines are a
family of parallel lines with slope $-1$ in the $\Delta _{a}$-$\Delta _{b}$
plane. The Hamiltonians with parameters at each line share the same ground
state with the same energy, which are determined only by the intercepts of
the line. The intercepts of the five representative lines are $-2$, $-1$, $0$
, $1$ and $2$. (c) Plot of the fidelity for the dynamic process
in the system with different time-dependent parameters in Eq. (\protect\ref%
{DE}). It indicates that when the time-dependent parameters only move along
a fixed line, the fidelity is always equal to $1$.}
\label{fig1}
\end{figure*}

\section{Model and fixed lines}

\label{Model and fixed line} We consider the following non-Hermitian
fermionic Hamiltonian $H=$ $H(\Delta _{a},\Delta _{b},\mu )$ on a lattice of
length $2N$ (even $N$)

\begin{eqnarray}
&&H=J\sum_{l=1}^{2N}c_{l}^{\dagger }c_{l+1}+\mathrm{H.c.}+\mu
\sum_{l=1}^{2N}\left( 1-2n_{l}\right)  \notag \\
&&+\sum_{j=1}^{N}(\Delta _{a}c_{2j}^{\dagger }c_{2j+1}^{\dagger }+\Delta
_{a}c_{2j}c_{2j-1}  \notag \\
&&+\Delta _{b}c_{2j+1}c_{2j}+\Delta _{b}c_{2j-1}^{\dagger }c_{2j}^{\dagger
}),  \label{H}
\end{eqnarray}%
where $c_{l}^{\dag }$ $(c_{l})$\ is a fermionic creation (annihilation)
operator on site $l$, $n_{l}=c_{l}^{\dag }c_{l}$, $J$ the tunneling rate,
real number $\Delta _{a}$\ the strength of the $p$-wave pair creation
(annihilation) on the odd (even) dimer while $\Delta _{b}$\ the strength of
the $p$-wave pair annihilation (creation) on the even (odd) dimer. $\mu $\
is the chemical potential. The non-Hermitian Kitaev model is
schematically illustrated in Fig. \ref{fig1}(a). For a closed chain, we
define $c_{2N+1}=c_{1}$ and for an open chain, we set $c_{2N+1}=0$. The
Kitaev model is known to have a rich phase diagram in its Hermitian version,
i.e.,\ $H\rightarrow H+H^{\dag }$, or $\Delta _{a}=\Delta _{b}$. In
particular, in the non-trivial topological region, the ground state is
two-fold degenerate in the large $N$\ limit, which stems from the Majorana
zero modes. In this work, we focus on what happens when $\Delta _{a}\neq
\Delta _{b}$. To be more precise, we investigate the influence of the
non-Hermitian terms on the topological phase and the edge modes of the
corresponding Majorana lattice.

We introduce the Fourier transformations in two sub-lattices

\begin{equation}
\left( 
\begin{array}{c}
c_{2j-1} \\ 
c_{2j}%
\end{array}%
\right) =\frac{1}{\sqrt{N}}\sum_{k}e^{ikj}\left( 
\begin{array}{c}
\alpha _{k} \\ 
\beta _{k}%
\end{array}%
\right) ,
\end{equation}%
and inversely, the spinless fermionic operators in $k$ space $\alpha _{k},$\ 
$\beta _{k}$ are%
\begin{equation}
\left( 
\begin{array}{c}
\alpha _{k} \\ 
\beta _{k}%
\end{array}%
\right) =\frac{1}{\sqrt{N}}\sum_{k}e^{-ikj}\left( 
\begin{array}{c}
c_{2j-1} \\ 
c_{2j}%
\end{array}%
\right)
\end{equation}%
where $j=1,2,...,N$, $k=2m\pi /N$, $m=0,1,2,...,N-1$. The Hamiltonian with
periodic boundary condition can be block diagonalized by this transformation
due to its translational symmetry, i.e.,%
\begin{equation}
H=\sum_{k\in \left[ 0,\pi \right] }H_{k}=H_{0}+H_{\pi }+\sum_{k\in (0,\pi
)}\psi _{k}^{\dagger }h_{k}\psi _{k},
\end{equation}%
satisfying $\left[ H_{k},H_{k^{\prime }}\right] =0$, where the operator
vector $\psi _{k}^{\dagger }=\left( 
\begin{array}{cccc}
\alpha _{k}^{\dagger } & \beta _{k}^{\dagger } & \alpha _{-k} & \beta _{-k}%
\end{array}%
\right) $, and the core matrix is expressed explicitly as

\begin{widetext}
\begin{equation}
	h_{k}=\left(
	\begin{array}{cccc}
		-2\mu & J\left( 1+e^{-ik}\right) & 0 & \Delta _{b}-\Delta _{a}e^{-ik} \\
		J\left( 1+e^{ik}\right) & -2\mu & \Delta _{a}e^{ik}-\Delta _{b} & 0 \\
		0 & \Delta _{b}e^{-ik}-\Delta _{a} & 2\mu & -J\left( 1+e^{-ik}\right) \\
		\Delta _{a}-\Delta _{b}e^{ik} & 0 & -J\left( 1+e^{ik}\right) & 2\mu%
	\end{array}%
	\right) .
\end{equation}%
\end{widetext}

Here, $H_{0}$ and $H_{\pi }$ have the form%
\begin{eqnarray}
H_{0} &=&2J\alpha _{0}^{\dagger }\beta _{0}+2J\beta _{0}^{\dagger }\alpha
_{0}+2\mu \left( \alpha _{0}\alpha _{0}^{\dagger }-\beta _{0}^{\dagger
}\beta _{0}\right)  \notag \\
&&+\left( \Delta _{b}-\Delta _{a}\right) \left( \alpha _{0}^{\dagger }\beta
_{0}^{\dagger }+\alpha _{0}\beta _{0}\right) , \\
H_{\pi } &=&2\mu \left( \alpha _{\pi }\alpha _{\pi }^{\dagger }-\beta _{\pi
}^{\dagger }\beta _{\pi }\right)  \notag \\
&&+\left( \Delta _{b}+\Delta _{a}\right) \left( \alpha _{\pi }^{\dagger
}\beta _{\pi }^{\dagger }+\beta _{\pi }\alpha _{\pi }\right) .
\end{eqnarray}%
The eigenvalue $\epsilon _{k}$\ and eigenvector $\left\vert \varphi
_{k}\right\rangle $\ of $h_{k}$, satisfying $h_{k}\left\vert \varphi
_{k}\right\rangle =\epsilon _{k}\left\vert \varphi _{k}\right\rangle $, can
be obtained analytically or numerically, which will be used for numerical
simulation of quench dynamics.

In the following, we focus on the case with $\mu =0$. The Hamiltonian $H$\
can be expressed in the diagonal form of the Hamiltonian 
\begin{equation}
H=H_{0}+H_{\pi }+\sum_{k\in (0,\pi )}\sum_{\rho \sigma }\varepsilon _{\rho
\sigma }^{k}\overline{A}_{\rho \sigma }^{k}A_{\rho \sigma }^{k}
\end{equation}%
where
\begin{eqnarray}
\varepsilon _{\rho \sigma }^{k} &=&\rho \sqrt{\left( 2J\cos \frac{k}{2}%
\right) ^{2}+\left( \Delta _{a}+\Delta _{b}\right) ^{2}\sin ^{2}\frac{k}{2}}
\notag \\
&&+i\sigma \left( \Delta _{a}-\Delta _{b}\right) \cos \frac{k}{2}. \label{ek}
\end{eqnarray}%
and the form of $A_{\rho \sigma }^{k}$\ is presented in the Appendix and
is independent of the value of\ $\left( \Delta _{a}-\Delta _{b}\right) $,
satisfying the canonical commutation relations%
\begin{eqnarray}
\{A_{\rho \sigma }^{k},\overline{A}_{\rho ^{\prime }\sigma ^{\prime
}}^{k^{\prime }}\} &=&\delta _{kk^{\prime }}\delta _{\rho \rho ^{\prime
}}\delta _{\sigma \sigma ^{\prime }},  \notag \\
\{A_{\rho \sigma }^{k},A_{\rho ^{\prime }\sigma ^{\prime }}^{k^{\prime }}\}
&=&\{\overline{A}_{\rho \sigma }^{k},\overline{A}_{\rho ^{\prime }\sigma
^{\prime }}^{k^{\prime }}\}=0.
\end{eqnarray}%
Then the ground state is%
\begin{eqnarray}
\left\vert \text{\textrm{G}}\right\rangle &=&\frac{1}{2}\left( \alpha _{\pi
}^{\dagger }\beta _{\pi }^{\dagger }+1\right) \left( \alpha _{0}^{\dagger
}-\beta _{0}^{\dagger }\right) \left\vert 0\right\rangle  \notag \\
&&\times \prod\limits_{k\in \left( 0,\pi \right) }\overline{A}_{-+}^{k}%
\overline{A}_{--}^{k}\left\vert \mathrm{Vac}\right\rangle
\end{eqnarray}%
with ground state energy%
\begin{equation}
E_{\mathrm{g}}=2\sum_{k\in (0,\pi )}\mathrm{Re}\varepsilon
_{-+}^{k}-(2J+\Delta _{a}+\Delta _{b}).
\end{equation}%
Here $\left\vert \mathrm{Vac}\right\rangle $\ is the vacuum state of the set
of operator \{$A_{\rho \sigma }^{k}$\}, i.e., $A_{\rho \sigma
}^{k}\left\vert \mathrm{Vac}\right\rangle =0$, while $\left\vert
0\right\rangle $\ is the vacuum state of the fermion operators. We find that 
$\left\vert \text{\textrm{G}}\right\rangle $\ and $E_{\mathrm{g}}$\ are both
independent of the strength of the non-Hermiticity $\Delta
_{a}-\Delta _{b}$ along the zero-$\mu $ line in the phase diagram. To describe this phenomenon, we dub the lines with slope $-1$ in
the $\Delta _{a}$-$\Delta _{b}$ plane "fixed lines", which are
clearly depicted in Fig. \ref{fig1}(b).

Based on the above analysis, one can consider a time-dependent Hamiltonian $%
H(t)$\ with the constraint d$(\Delta _{a}+\Delta _{b})/$d$t=0$\ but d$%
(\Delta _{a}-\Delta _{b})/$d$t\neq 0$. Considering the fact that the
groundstate and the energy are independent of the value of $\Delta _{a}-\Delta
_{b}$, when d$(\Delta _{a}+\Delta _{b})/$d$t=0$, based on the
fact 
\begin{equation}
H(t)\left\vert \text{\textrm{G}}\right\rangle =H(t^{\prime })\left\vert 
\text{\textrm{G}}\right\rangle =\varepsilon _{g}\left\vert \text{\textrm{G}}%
\right\rangle ,
\end{equation}
or
\begin{equation}
\exp [-iH(n\tau )\tau ]\left\vert \text{\textrm{G}}\right\rangle
=e^{i\varepsilon _{g}\tau }\left\vert \text{\textrm{G}}\right\rangle ,
\end{equation}%
we have
\begin{eqnarray}
\left\vert \Phi (t)\right\rangle &=&\mathcal{T}\exp [-i\int_{0}^{t}H(t)\text{%
d}t]\left\vert \text{\textrm{G}}\right\rangle  \notag \\
&=&\lim_{\tau \rightarrow 0}\mathcal{T}\prod_{n=1}^{N}\exp \left(
-i\varepsilon _{g}\tau \right) \left\vert \text{ \textrm{G}}\right\rangle 
\notag \\
&=&e^{i\phi }\left\vert \text{\textrm{G}}\right\rangle,
\end{eqnarray}%
where $\mathcal{T}$ is the time-order operator and $\phi $\ is an overall
phase. The Hamiltonian $H(t)$ is the Hamiltonian shown in Eq.
(1) with time-dependent parameters $\Delta _{a}(t)-\Delta _{b}(t)$. In Fig. \ref{fig1}(c), we provide the fidelity 
\begin{equation}
F\left( t\right) =|\left\langle \Phi (0)\right\vert \Phi (t)\rangle |^{2},
\label{fidelity}
\end{equation}%
for a concrete system with 
\begin{equation}
\Delta _{a}=1+\sin t^{2},\Delta _{b}=1-\sin \left( t^{2}+\zeta \right) ,
\label{DE}
\end{equation}%
by taking different $\zeta =0$, $0.5$, $1$. The plot shows that the fidelity
is always equal to $1$ at the fixed line. This allows the constancy of the
winding number extracted from the ground state in the process of varying the
balance strength at an arbitrary rate d$(\Delta _{a}-\Delta _{b})/$d$t$,
exhibiting the robustness of the topology for non-Hermitian Kitaev chain
under time-dependent perturbations d$(\Delta _{a}-\Delta _{b})/$d$t$ \cite{SYBPRB2}. However, the situation may change under the open
condition, which will be investigated in the following section.

\begin{figure}[tbh]
\centering\includegraphics[width=0.48\textwidth]{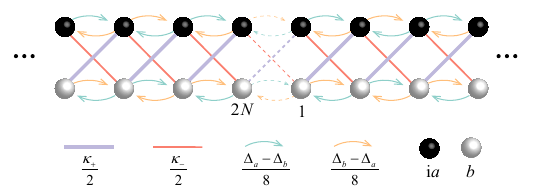} 
\caption{Geometry for the Majorana lattice described in Eq. (\protect\ref{Mr}). The system consists of two sublattices, $ia$ and $b$, indicated by black
and white circles, respectively. The red and purple solid lines indicate the
couplings between the sublattices.\ The yellow and green arrows are the
unidirectional hopping. There is an engineered impurity, which is labeled as
the dashed lines and arrows. The corresponding zero modes are plotted in
Fig. (\protect\ref{fig3}).} \label{fig2}
\end{figure}

\section{Majorana zero modes}

\label{Majorana zero modes} As mentioned above, it has been appreciated
previously that bulk-boundary correspondence holds in the Hermitian Kitaev
model. The analysis in last section shows that the ground state remains
unchanged at the fixed lines, maintaining the topology feature in the
presence of non-Hermitian terms. However, the periodic condition plays an
important role in such a conclusion, which will also be demonstrated in the
next section in the context of the quantum spin model. In other words, the
non-Hermitian terms should break the constancy of the ground state when an
open boundary condition is taken. A natural question is whether edge modes
still exist for the opened Majorana lattice in the presence of non-Hermitian
terms.

We begin with a defected Kitaev system by introducing a resonant impurity
located at a dimer across two sites $\left( 1,2N\right) $. The Hamiltonians
reads%
\begin{eqnarray}
H_{\mathrm{D}} &=&H-\lambda M,  \notag \\
M &=&\Delta _{a}c_{2N}^{\dagger }c_{1}^{\dagger }+\Delta
_{b}c_{1}c_{2N}+Jc_{2N}^{\dagger }c_{1}+Jc_{1}^{\dagger }c_{2N},  \label{H_D}
\end{eqnarray}%
where $\lambda $\ controls the strength of the impurity and $\lambda =1$\
indicates the open chain. We introduce Majorana fermion operators%
\begin{equation}
a_{j}=c_{j}^{\dagger }+c_{j},b_{j}=-i\left( c_{j}^{\dagger }-c_{j}\right) ,
\label{ab}
\end{equation}%
which satisfy the relations%
\begin{eqnarray}
\left\{ a_{j},a_{j^{\prime }}\right\} &=&2\delta _{j,j^{\prime }},\left\{
b_{j},b_{j^{\prime }}\right\} =2\delta _{j,j^{\prime }},  \notag \\
\left\{ a_{j},b_{j^{\prime }}\right\} &=&0.
\end{eqnarray}%
The inverse transformation is 
\begin{equation}
c_{j}^{\dagger }=\frac{1}{2}\left( a_{j}+ib_{j}\right) ,c_{j}=\frac{1}{2}%
\left( a_{j}-ib_{j}\right) .
\end{equation}

The Majorana representation of the Hamiltonian has the form

\begin{eqnarray}
&&H_{\mathrm{D}}=\sum_{l=1}^{N}[\kappa _{+}\left(
ib_{2j}a_{2j+1}+ib_{2j-1}a_{2j}\right)  \notag \\
&&+\kappa _{-}\left( ib_{2j}a_{2j-1}+ib_{2j+1}a_{2j}\right) +\frac{\Delta
_{a}-\Delta _{b}}{4}  \notag \\
&&\times \left(
a_{2j}a_{2j+1}+b_{2j+1}b_{2j}+a_{2j}a_{2j-1}+b_{2j-1}b_{2j}\right) ]  \notag
\\
&&-\lambda \lbrack \kappa _{+}ib_{2N}a_{1}+\kappa _{-}ib_{1}a_{2N}+\frac{%
\Delta _{a}-\Delta _{b}}{4}  \notag \\
&&\times (a_{2N}a_{1}+b_{1}b_{2N})],
\end{eqnarray}%
with $\kappa _{\pm }=[2J\pm (\Delta _{a}+\Delta _{b})]/4$. We write down the
Hamiltonian in the basis $\varphi ^{\dagger }=(-ia_{1},$ $b_{1},$ $-ia_{2},$ 
$b_{2},$ $-ia_{3},$ $b_{3},$ $...)$ in the form%
\begin{equation}
H_{\mathrm{D}}=\varphi ^{\dagger }h_{\mathrm{D}}\varphi ,
\end{equation}%
where $h_{\mathrm{D}}$\ represents a $4N\times 4N$ matrix. Here, the matrix $%
h_{\mathrm{D}}$\ can be explicitly written as 
\begin{widetext}
\begin{eqnarray}
	&&h_{\mathrm{D}}=\frac{\Delta _{a}-\Delta _{b}}{8}\sum_{j=1}^{N}(\left\vert
	2j,A\right\rangle \left\langle 2j+1,A\right\vert +\left\vert
	2j+1,B\right\rangle \left\langle 2j,B\right\vert +\left\vert
	2j,A\right\rangle \left\langle 2j-1,A\right\vert +\left\vert
	2j-1,B\right\rangle \left\langle 2j,B\right\vert )-\mathrm{H.c.}  \notag \\
	&&+\sum_{l=1}^{2N}(\frac{\kappa _{+}}{2}\left\vert l,B\right\rangle \left\langle
	l+1,A\right\vert +\frac{\kappa _{-}}{2}\left\vert l+1,B\right\rangle \left\langle
	l,A\right\vert +\mathrm{H.c.})  \notag \\
	&&-\lambda \lbrack \frac{\kappa _{+}}{2}(\left\vert 2N,B\right\rangle \left\langle
	1,A\right\vert +\mathrm{H.c.})+\frac{\kappa _{-}}{2}(\left\vert 1,B\right\rangle
	\left\langle 2N,A\right\vert +\mathrm{H.c.})+\frac{\Delta _{a}-\Delta _{b}}{8%
	}(\left\vert 2N,A\right\rangle \left\langle 1,A\right\vert +\left\vert
	1,B\right\rangle \left\langle 2N,B\right\vert -\mathrm{H.c.})].\label{Mr}
\end{eqnarray}
\end{widetext}

In Fig. \ref{fig2} the geometry of the lattice $h_{\mathrm{D}}$\ is
illustrated. We start with the perfect case with $\lambda =0$, and the
matrix describes a uniform ladder with unequal hopping strength\ under the
periodic boundary condition. The spectrum of the matrix can be easily
obtained as 

\begin{eqnarray}
\mathcal{E}_{K} &=&\pm \frac{1}{4}\sqrt{\left( 2J\right) ^{2}\cos
^{2}K+\left( \Delta _{a}+\Delta _{b}\right) ^{2}\sin ^{2}K}  \notag \\
&&\pm i \frac{\left( \Delta _{a}-\Delta _{b}\right) }{4}\cos
K ,
\end{eqnarray}
which is one quarter of $\varepsilon _{\rho \sigma }^{k}$ in Eq. (\ref{ek}),
where the wave vector $K=m\pi /N$, $m=0,1,2,...,$ $N-1$. When $%
\mathcal{E}_{K}$\ is considered\ as the energy band of a non-Hermitian
system, i.e., a tight-binding ladder,\ the\ half-filled groundstate energy
can be obtained as%
\begin{equation}
E_{\text{g}}=-\frac{N}{\pi }\max \left( 2J,\Delta _{a}+\Delta _{b}\right) 
\mathrm{E}\left( e\right)
\end{equation}%
which is obviously independent of $\Delta _{a}-\Delta _{b}$. Here function $%
E $\ means the complete elliptic integral of the second kind and $e$\
represents the eccentricity of the ellipse with width $2J$\ and height $%
\Delta _{a}+\Delta _{b}$.

\begin{figure}[t]
\centering\includegraphics[width=0.45\textwidth]{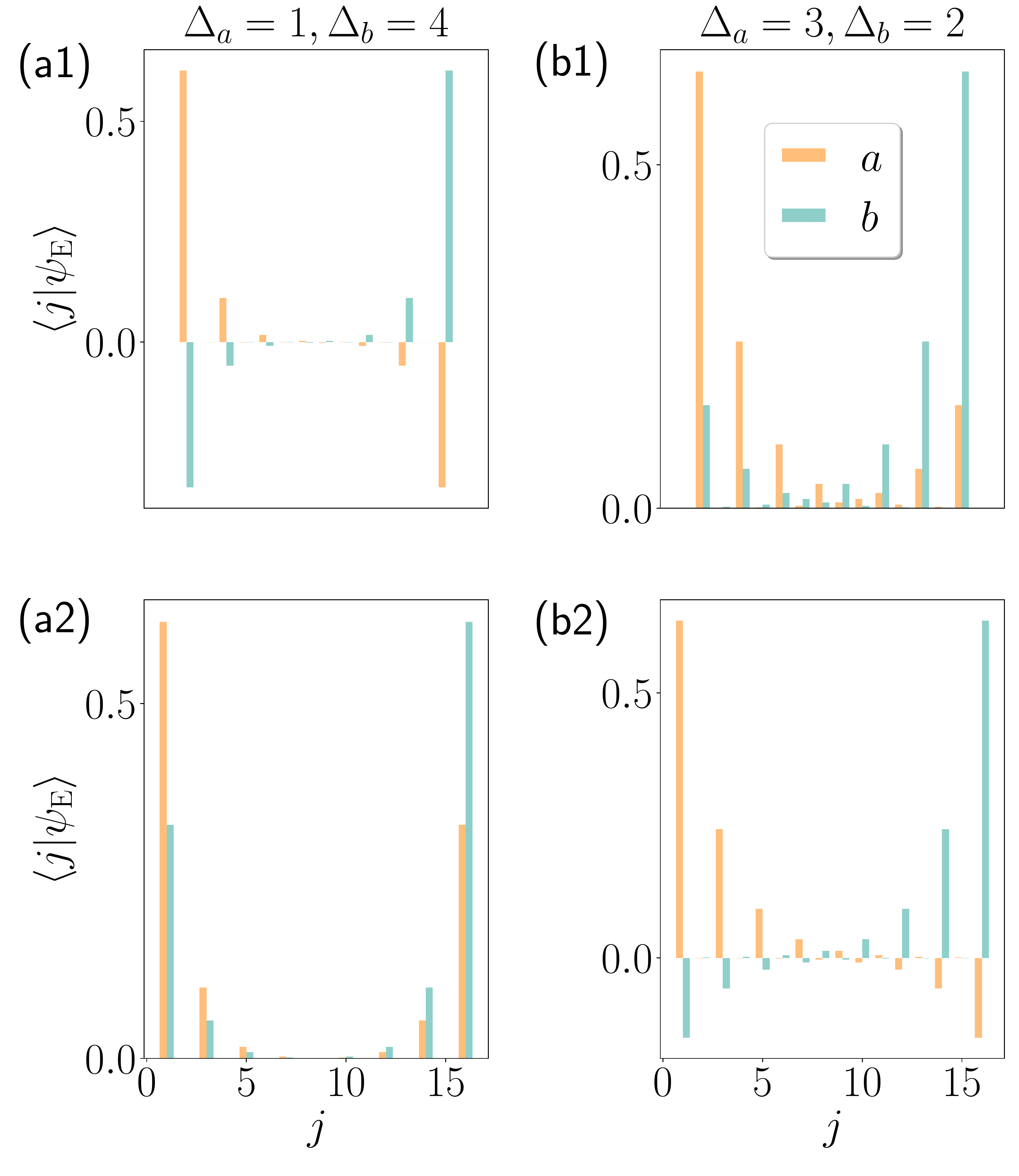} 
\caption{The profiles of the zero modes from Eq. \protect\ref{edge state}
for Majorana lattices in Fig. \protect\ref{fig2} with $N=8$. The first row
is the plot for the system with the condition $\protect\lambda =\protect\lambda _{-}$ and the second row with the condition $\protect\lambda =\protect\lambda _{+}$. The color bars indicate the amplitudes of two
different sublattices. All the four patterns exhibit evident skin effect.} %
\label{fig3}
\end{figure}

The translational symmetry is broken when we take $\lambda \neq 0$. Our goal
is to find out the zero modes when taking the open boundary.\ To this end,
we consider the cases with resonant conditions $\lambda =\lambda
_{+}=1-\gamma ^{-N}$\ and $\lambda _{-}=1-\gamma ^{N}$, respectively. Under this condition, there always exist two zero modes for any
given $N$. Importantly, it provides explicit expressions of the zero
modes even in the finite $N$\ limit. The coefficient $\gamma $\ has the form%
\begin{equation}
\gamma =\frac{\left\vert \Delta _{a}+\Delta _{b}\right\vert \sqrt{\left(
\Delta _{a}-\Delta _{b}\right) ^{2}+4J^{2}}-2J^{2}-\Delta _{a}^{2}-\Delta
_{b}^{2}}{2\left( J^{2}-\Delta _{a}\Delta _{b}\right) },
\end{equation}%
which determines the values of $\lambda =1$\ or $\infty $,\ in\ the large $N$%
\ limit. Accordingly, $H_{\mathrm{D}}$\ describes an open ladder of $4N$ and 
$\left( 4N-4\right) $ sites respectively, when $N$ turns to infinity.
Straightforward derivations show the following exact results. There are a
pair of zero modes%
\begin{equation}
h_{\mathrm{D}}\left\vert \psi _{\mathrm{L}}\right\rangle =h_{\mathrm{D}%
}\left\vert \psi _{\mathrm{R}}\right\rangle =0,
\end{equation}%
for arbitrary $N$, when taking $\lambda =\lambda _{\pm }$. Leaving aside the
overall normalization, (i) for $\lambda =\lambda _{-}$, we have the explicit
expression of zero modes%
\begin{equation}
\left\{ 
\begin{array}{l}
\left\vert \psi _{\mathrm{L}}\right\rangle =\sum\limits_{j=1}^{2N}\gamma
^{j-1}(\left\vert 2j-1,A\right\rangle +\beta \left\vert 2j-1,B\right\rangle )
\\ 
\left\vert \psi _{\mathrm{R}}\right\rangle =\sum\limits_{j=1}^{2N}\gamma
^{N-j}(\left\vert 2j,B\right\rangle +\beta \left\vert 2j,A\right\rangle )%
\end{array}%
\right. ,  \label{lam_-}
\end{equation}%
while (ii) for $\lambda =\lambda _{+}$, we have%
\begin{equation}
\left\{ 
\begin{array}{l}
\left\vert \psi _{\mathrm{R}}\right\rangle =\sum\limits_{j=1}^{2N}\gamma
^{N-j}(\left\vert 2j-1,B\right\rangle -\beta \left\vert 2j-1,A\right\rangle )
\\ 
\left\vert \psi _{\mathrm{L}}\right\rangle =\sum\limits_{j=1}^{2N}\gamma
^{j-1}(\left\vert 2j,A\right\rangle -\beta \left\vert 2j,B\right\rangle )%
\end{array}%
\right. ,  \label{lam_+}
\end{equation}%
where the coefficient is $\beta =[\sqrt{4J^{2}+(\Delta _{b}-\Delta _{a})^{2}}
$ $-2J]/(\Delta _{b}-\Delta _{a})$. The exact result is demonstrated by the
plot of the profile of the normalized edge state 
\begin{equation}
\left\vert \psi _{\mathrm{E}}\right\rangle =\frac{\left\vert \psi _{\mathrm{R%
}}\right\rangle }{\sqrt{2}\left\vert \left\vert \psi _{\mathrm{R}%
}\right\rangle \right\vert }+\frac{\left\vert \psi _{\mathrm{L}%
}\right\rangle }{\sqrt{2}\left\vert \left\vert \psi _{\mathrm{L}%
}\right\rangle \right\vert },  \label{edge state}
\end{equation}%
\ in Fig. \ref{fig3}\ for finite $N$. It is obvious that the zero modes are
standard edge modes, demonstrating the bulk-edge correspondence at $\lambda
=1$\ in the thermodynamic limit.

Compared to the ground state for periodic boundary condition, the profiles
of edge modes $\left\vert \psi _{\mathrm{L,R}}\right\rangle $ are\ now
dependent on the value of $\Delta _{a}-\Delta _{b}$. This means that the
non-Hermitian components have an effect on the eigenstates when the
translational symmetry is broken. It may be because the balance between the
staggered imbalanced pair creation and annihilation terms is broken. To
understand how the non-Hermitian components cancel each other out, in the
following section, we study the model in the spin representation.

\section{Dissipationless ferromagnetic order}

\label{Dissipationless ferromagnetic order} It has been noted that the
Kitaev chain, a $p$-wave superconductor can be mapped into a quantum spin $%
XY $\ model with a transverse field via a Jordan-Wigner transformation \cite%
{Pfeuty},\ replacing the spinless fermion creation and annihilation
operators with spin flip operators. Accordingly, the present non-Hermitian
Kitaev model should correspond to a non-Hermitian $XY$\ model. Intuitively,
a spin model can provide a clear physical picture for understanding the
balance of non-Hermitian terms from different locations. The non-Hermitian
extensions for quantum spin $XY$\ model have been studied via several
examples \cite{ZXZPRB,LCPRA90,LCPRA94}. However, the non-Hermitian spin
model mapped from a Kitaev model is slightly special. It possesses a subtle
boundary term involving a string operator. In this section, we consider this
problem based on the simplest case with $\Delta _{a}+\Delta _{b}=2J$.

Introducing the Jordan Winger transformation%
\begin{equation}
c_{j}=\prod_{i=1}^{j-1}\sigma _{i}^{z}\sigma _{j}^{-},c_{j}^{\dagger
}=\prod_{i=1}^{j-1}\sigma _{i}^{z}\sigma _{j}^{+},
\end{equation}%
the Kitaev Hamiltonian can be expressed as 
\begin{equation}
H=JH_{0}+\frac{i\left( \Delta _{a}-\Delta _{b}\right) }{4}\mathcal{H},
\end{equation}%
where both terms%
\begin{equation}
H_{0}=-\sum_{l=1}^{2N-1}\sigma _{l}^{x}\sigma _{l+1}^{x}-\left(
\prod\limits_{i=2}^{2N-1}\sigma _{i}^{z}\right) \sigma _{1}^{y}\sigma
_{2N}^{y},  \label{H_0}
\end{equation}%
and 
\begin{eqnarray}
&&\mathcal{H}=\sum_{j=1}^{N}(\sigma _{2j-1}^{x}\sigma _{2j}^{y}+\sigma
_{2j-1}^{y}\sigma _{2j}^{x})  \notag \\
&&-\sum_{j=1}^{N-1}(\sigma _{2j}^{y}\sigma _{2j+1}^{x}+\sigma
_{2j}^{x}\sigma _{2j+1}^{y})  \notag \\
&&+\left( \prod\limits_{i=2}^{2N-1}\sigma _{i}^{z}\right) \left( \sigma
_{1}^{y}\sigma _{2N}^{x}+\sigma _{2N}^{y}\sigma _{1}^{x}\right) ,
\end{eqnarray}%
are all Hermitian. We note that $H_{0}$\ represents a simplest Ising model
but with a subtle boundary condition $\left( \Pi _{i=2}^{2N-1}\sigma
_{i}^{z}\right) \sigma _{1}^{y}\sigma _{2N}^{y}$, which contains a
string-like nonlocal operator $\Pi _{i=1}^{2N}\sigma _{i}^{z}$. The effect
of the boundary condition has rarely been considered in previous
investigations on the Ising model and is important in this work. It is well
known that the GHZ states \cite{DMG,DB,PJW}%
\begin{equation}
\left\vert \mathrm{GHZ}^{\pm }\right\rangle =\frac{1}{\sqrt{2}}\left(
\prod_{i=1}^{2N}\left\vert \uparrow \right\rangle _{i}\pm
\prod_{i=1}^{2N}\left\vert \downarrow \right\rangle _{i}\right) ,
\end{equation}%
\ are the ground state doublet of $H_{0}$\ when the boundary condition is
modified by taking $\left( \Pi _{i=2}^{2N-1}\sigma _{i}^{z}\right) \sigma
_{1}^{y}\sigma _{2N}^{y}$ $\longrightarrow \sigma _{1}^{x}\sigma _{2N}^{x}$\
or $\left( \Pi _{i=2}^{2N-1}\sigma _{i}^{z}\right) \sigma _{1}^{y}\sigma
_{2N}^{y}$ $\longrightarrow 0$. Here the kets are defined as $\sigma
_{i}^{x}\left\vert \uparrow \right\rangle _{i}=\left\vert \uparrow
\right\rangle _{i}$\ and $\sigma _{i}^{x}\left\vert \downarrow \right\rangle
_{i}=-\left\vert \downarrow \right\rangle _{i}$. However, straightforward
derivation shows that 
\begin{equation}
H_{0}\left\vert \mathrm{GHZ}^{\pm }\right\rangle =\left( -2N+1\pm 1\right)
\left\vert \mathrm{GHZ}^{\pm }\right\rangle ,
\end{equation}%
which indicates that $\left\vert \mathrm{GHZ}^{-}\right\rangle $\ is the
ground state singlet of $H_{0}$.\ Remarkably, when $\mathcal{H}$
is taken into account, one can find the relation\ 
\begin{equation}
\sigma _{l}^{y}(\sigma _{l-1}^{x}-\sigma _{l+1}^{x})\left\vert \mathrm{GHZ}%
^{\pm }\right\rangle =0,
\end{equation}%
for $l\in \left[ 2,2N-1\right] $ in the bulk. It is clear that the
non-Hermitian spin coupling terms from two\ neighboring dimers cancel each
other out when they are applied to the bulk of the ferromagnetic state.
Furthermore, applying $\mathcal{H}$ to the state $\left\vert 
\mathrm{GHZ}^{\pm }\right\rangle $, we have%
\begin{eqnarray}
&&\mathcal{H}\left\vert \mathrm{GHZ}^{\pm }\right\rangle =(\sigma
_{1}^{y}\sigma _{2}^{x}+\sigma _{2N-1}^{x}\sigma _{2N}^{y})\left\vert 
\mathrm{GHZ}^{\pm }\right\rangle  \notag \\
&&+\left( \sigma _{1}^{y}\sigma _{2N}^{x}+\sigma _{2N}^{y}\sigma
_{1}^{x}\right) \left( \prod_{i=2}^{2N-1}\sigma _{i}^{z}\right) \left\vert 
\mathrm{GHZ}^{\pm }\right\rangle ,
\end{eqnarray}%
i.e., the action of $\mathcal{H}$ on state $\left\vert \mathrm{%
GHZ}^{\pm }\right\rangle $ is only left at the boundary. Then we have%
\begin{equation}
\mathcal{H}\left\vert \mathrm{GHZ}^{-}\right\rangle =0,
\end{equation}%
due to the fact%
\begin{eqnarray}
&&\left( \sigma _{1}^{y}\sigma _{2N}^{x}+\sigma _{2N}^{y}\sigma
_{1}^{x}\right) \left( \prod_{i=2}^{2N-1}\sigma _{i}^{z}\right) \left\vert 
\mathrm{GHZ}^{\pm }\right\rangle  \notag \\
&=&\left( \sigma _{1}^{x}\sigma _{2N}^{y}+\sigma _{2N}^{x}\sigma
_{1}^{y}\right) \left( \pm \right) \left\vert \mathrm{GHZ}^{\pm
}\right\rangle .
\end{eqnarray}

The above investigation has shown that $\left\vert \mathrm{GHZ}%
^{-}\right\rangle $ is still the ground state of $H$. The groundstate wave
function is independent of the value of $\left( \Delta _{a}-\Delta
_{b}\right) $ which accords with the analysis in last section. The
underlying mechanism is that a pair of non-Hermitian terms from two
neighboring dimers cancel each other out, leaving the GHZ states with
perfect ferromagnetic order. For the other phases, $\Delta
_{a}+\Delta _{b}\ne2J$, next to the fixed line we mentioned in this section,
there also exists a corresponding fixed line on which the ground state is
the same as that of the corresponding Hermitian quantum XY model. The
topological feature of the Hamiltonian with $\Delta_a\ne0,\Delta_b=0$ has
been studied in Ref. \cite{SYBPRB2}. In addition, the obtained result\ can
be applied to other quantum spin systems. For example, when we consider a
Heisenberg ring with an additional non-Hermitian term $i\sum_{j=1}^{N}(%
\sigma _{2j-1}^{x}\sigma _{2j}^{y}-\sigma _{2j}^{y}\sigma _{2j+1}^{x})$, it
is easy to find that the ferromagnetic states in the $x$ direction remain
unchanged. The ferromagnetic order does not dissipate in the presence of
such non-Hermitian fluctuations.

\begin{figure*}[tbh]
\centering \includegraphics[width=0.9\textwidth]{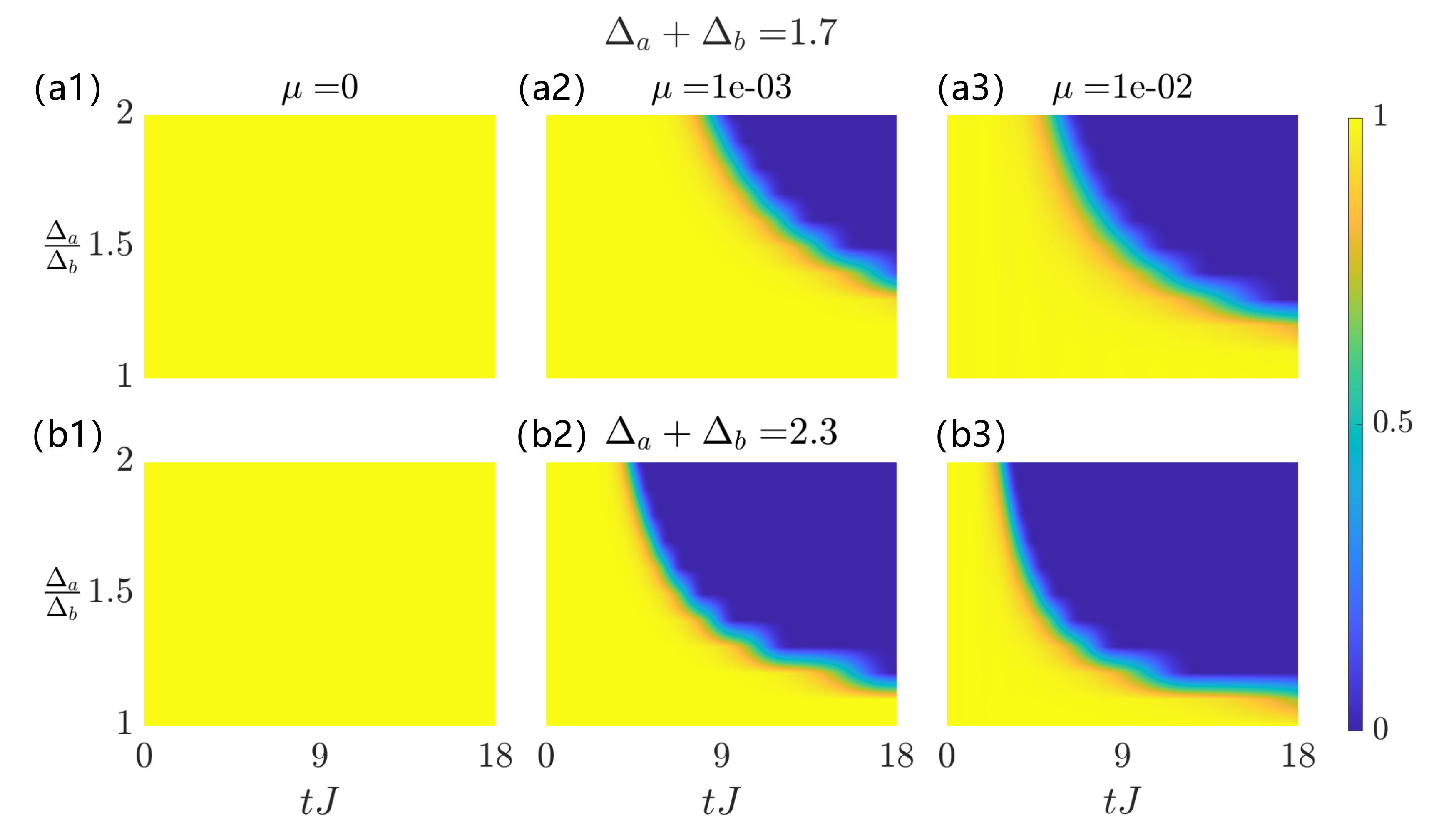}
\caption{$2$D color contour plots of the fidelity in Eq. (\protect\ref{Ft})
for the quench processes. The system parameters $\Delta _{a}/\Delta _{b}$\
and $\protect\mu $\ are indicated in the panels. The unitary fidelity in
(a1) and (b1) demonstrates our exact results. (a2, a3) and (b2, b3) are the
cases with two different values of small $\protect\mu $. The deep shadows
indicate the region of zero fidelity. The sharp edge indicates the sudden
drops of the fidelity.}
\label{fig4}
\end{figure*}

\section{Stability of quench dynamics}

\label{Stability of quench dynamics} In this section, we present the
features of the dynamic behavior when the quench Hamiltonian deviates from
the fixed lines by small nonzero $\mu $. To be precise, in the following, we
consider the numerical simulation of the quench process under the
Hamiltonian 
\begin{equation}
H_{\text{\textrm{Pre}}}=H(\Delta _{a},\Delta _{b},0),H_{\text{\textrm{Pos}}%
}=H(\Delta _{a},\Delta _{b},\mu ),
\end{equation}%
where $H(\Delta _{a},\Delta _{b},\mu )$ is defined at the beginning in Eq. (%
\ref{H}). To capture the effect of small $\mu $ on the dynamics, we
introduce the concept of fidelity, which is a measure of the stability of
the original quantum state under the perturbation. An initial quantum state $%
\left\vert \Phi (0)\right\rangle $ evolves during time $t$ under postquench
Hamiltonian $H_{\text{\textrm{Pos}}}$ reaching state $\left\vert \Phi
(t)\right\rangle =e^{-iH_{\text{\textrm{Pos}}}t}\left\vert \Phi
(0)\right\rangle $. The fidelity is defined as%
\begin{equation}
F\left( t\right) =|\left\langle \Phi (0)\right\vert e^{-iH_{\text{\textrm{Pos%
}}}t}\left\vert \Phi (0)\right\rangle |^{2}.  \label{Ft}
\end{equation}%
In our study, the initial state is prepared as the ground state of $H_{\text{%
\textrm{Pre}}}$, i.e., $\left\vert \Phi (0)\right\rangle =\left\vert \mathrm{%
G}\right\rangle $. Obviously, we always have $F\left( t\right) =1$\ when $H_{%
\text{\textrm{Pre}}}=H(\Delta _{a},\Delta _{b},0)$ and $H_{\text{\textrm{Pos}%
}}=H(\Delta _{a}^{\prime },\Delta _{b}^{\prime },0)$\ with $\Delta
_{a}+\Delta _{b}=\Delta _{a}^{\prime }+\Delta _{b}^{\prime }$.\ It is
expected that the fidelity obeys $F\left( t\right) \approx 1$ within a
period of time in the presence small $\mu $. The numerical results of $%
F\left( t\right) $ obtained by exact diagonalization of each invariant
subspace indicated by $k$.

The results for systems with representative parameters $\Delta _{a}+\Delta
_{b}>2J$ and $<2J$ are presented in Fig. \ref{fig4}. We can see that the
results are in accord with our predictions for zero $\mu $. We note that the
fidelity experiences a sudden drop from $1$ to $0$ when the time reaches a
boundary. For a fixed small $\mu $, such a boundary in the time axis shrinks
as the ratio $\Delta _{a}/\Delta _{b}$\ increases. On the other hand, for a
fixed ratio $\Delta _{a}/\Delta _{b}$, such a boundary in the time axis
shrinks as the $\mu $\ increases. The aim of Fig. \ref{fig4} is
to demonstrate the continuity of the ground state when the parameter $\mu $
crosses over the zero point. The result indicates that there is no sudden
change in the ground state when the parameter deviates from the fixed
lines, i.e., each fixed line is not the critical line or quantum phase
boundary.

\section{Summary}

\label{summary}

In summary, we investigate the stability of the ground state of the p-wave
Kitaev model at fixed lines in the phase diagram in the presence of a
non-Hermitian pair term. Based on the exact solutions, we find that the
ground state remains unchanged in the presence of a restrained non-Hermitian
pair term under the periodic boundary condition for a finite system. When
the translational symmetry is broken, the non-Hermitian term affects on the
ground state for a finite system. However, the exact solution shows that a
resonant non-Hermitian impurity can induce a pair of zero modes in the
corresponding Majorana lattice, which asymptotically approach to the edge
modes in the thermodynamic limit, manifesting the bulk-boundary
correspondence. We have investigated the underlying mechanism through the
equivalent quantum spin system obtained by the Jordan-Wigner transformation
for finite chains. The numerical simulation of quench dynamics shows that
the obtained results still hold for slight deviation from the fixed lines. The main novelty of the present work is that we proposed a
non-Hermitian Kitaev model, with the non-Hermiticity arising from locally
imbalanced but globally balanced pair creation and annihilation. To date, the
proposed non-Hermitian Kiteav model in the literature mainly comes from the
imaginary potentials. This study provides the insight into the topological
phase that emerges from the interplay between spatially separated pairing
processes. In addition, the corresponding Majorana lattice of our model is
non-Hermitian but possesses zero edge modes. This model can be realized in
photonic systems, and the edge modes can be detected in experiments \cite%
{SWeiman}. Recently, it has been shown that the phase diagram of a Hermitian
Kitaev model can be demonstrated in the dynamic process \cite{SYBPRB}. This
will be an interesting topic for our future work to investigate the dynamics
of the present model.

\appendix

\section{Appendix}

In this appendix, we derive the solution of the Hamiltonian $H=H_{0}+H_{\pi
}+\sum_{k\in (0,\pi )}\psi _{k}^{\dagger }h_{k}\psi _{k}$\ with zero $\mu $,
where $\psi _{k}^{\dagger }=\left( 
\begin{array}{cccc}
\alpha _{k}^{\dagger } & \beta _{k}^{\dagger } & \alpha _{-k} & \beta _{-k}%
\end{array}%
\right) $ and\ the core matrix $h_{k}$\ can be rewritten in the form 
\begin{equation}
h_{k}=J\Gamma _{1}+\frac{\Delta _{a}+\Delta _{b}}{2}\Gamma _{2}+\frac{\Delta
_{a}-\Delta _{b}}{2}\Gamma _{3}.
\end{equation}%
Here, matrices 
\begin{equation}
\Gamma _{1}=\Gamma _{1}^{\dag }=\left( 
\begin{array}{cccc}
0 & \gamma _{-k} & 0 & 0 \\ 
\gamma _{k} & 0 & 0 & 0 \\ 
0 & 0 & 0 & -\gamma _{-k} \\ 
0 & 0 & -\gamma _{k} & 0%
\end{array}%
\right) ,
\end{equation}%
and%
\begin{equation}
\Gamma _{2}=\Gamma _{2}^{\dag }=\left( 
\begin{array}{cccc}
0 & 0 & 0 & \gamma _{\pi -k} \\ 
0 & 0 & -\gamma _{\pi +k} & 0 \\ 
0 & -\gamma _{\pi -k} & 0 & 0 \\ 
\gamma _{\pi +k} & 0 & 0 & 0%
\end{array}%
\right) ,
\end{equation}%
\ while%
\begin{equation}
\Gamma _{3}=-\Gamma _{3}^{\dag }=\left( 
\begin{array}{cccc}
0 & 0 & 0 & -\gamma _{-k} \\ 
0 & 0 & \gamma _{k} & 0 \\ 
0 & -\gamma _{-k} & 0 & 0 \\ 
\gamma _{k} & 0 & 0 & 0%
\end{array}%
\right) ,
\end{equation}%
is anti-Hermitian with $\gamma _{k}=1+e^{ik}$. Obviously, we have%
\begin{equation}
\left[ h_{k}\left( \Delta _{a},\Delta _{b}\right) \right] ^{\dag
}=h_{k}\left( \Delta _{b},\Delta _{a}\right) .
\end{equation}%
Direct derivation shows that%
\begin{equation}
H_{k}=\sum_{\rho \sigma }\varepsilon _{\rho \sigma }^{k}\overline{A}_{\rho
\sigma }^{k}A_{\rho \sigma }^{k},
\end{equation}%
for $k\in (0,\pi )$\ where 
\begin{eqnarray}
A_{\rho \sigma }^{k} &=&\frac{1}{\sqrt{\Omega }}[(1+\rho \sigma ie^{(i\theta
-k/2)})\alpha _{k}  \notag \\
&&+(\rho e^{i\left( \theta -k\right) }+i\sigma e^{-ik/2})\beta _{k}  \notag
\\
&&+(1-\rho \sigma ie^{i\left( \theta -k/2\right) })\alpha _{-k}^{\dagger } 
\notag \\
&&+(-\rho e^{i\left( \theta -k\right) }+i\sigma e^{-ik/2})\beta
_{-k}^{\dagger }],
\end{eqnarray}%
and%
\begin{equation}
\overline{A}_{\rho \sigma }^{k}=\left( A_{\rho \sigma }^{k}\right) ^{\dag
}(\rho \rightarrow -\rho ,\sigma \rightarrow -\sigma ),
\end{equation}%
with\ 
\begin{equation}
\tan \theta =\frac{\sin k\left( 2J-\Delta _{a}-\Delta _{b}\right) }{%
2J+\Delta _{a}+\Delta _{b}+\cos k\left( 2J-\Delta _{a}-\Delta _{b}\right) },
\end{equation}%
\ and $\sqrt{\Omega }$\ being the normalization factors. The spectrum is
\begin{eqnarray}
\varepsilon _{\rho \sigma }^{k} &=&\rho \sqrt{\left( 2J\cos \frac{k}{2}%
\right) ^{2}+\left( \Delta _{a}+\Delta _{b}\right) ^{2}\sin ^{2}\frac{k}{2}}
\notag \\
&&+i\sigma \left( \Delta _{a}-\Delta _{b}\right) \cos \frac{k}{2}.
\end{eqnarray}%
 The Hamiltonian $H_{k}$\ is diagonalized since the set of operators ($%
\overline{A}_{\rho \sigma }^{k},A_{\rho \sigma }^{k}$)\ satisfies the
canonical commutation relations%
\begin{eqnarray}
\{A_{\rho \sigma }^{k},\overline{A}_{\rho ^{\prime }\sigma ^{\prime
}}^{k^{\prime }}\} &=&\delta _{kk^{\prime }}\delta _{\rho \rho ^{\prime
}}\delta _{\sigma \sigma ^{\prime }},  \notag \\
\{A_{\rho \sigma }^{k},A_{\rho ^{\prime }\sigma ^{\prime }}^{k^{\prime }}\}
&=&\{\overline{A}_{\rho \sigma }^{k},\overline{A}_{\rho ^{\prime }\sigma
^{\prime }}^{k^{\prime }}\}=0.
\end{eqnarray}%
Notably, we have 
\begin{eqnarray}
&&A_{\rho \sigma }^{k}A_{\rho \overline{\sigma }}^{k}=(\overline{A}_{\rho 
\overline{\sigma }}^{k}\overline{A}_{\rho \sigma }^{k})^{\dag }  \notag \\
&=&\frac{1}{\Omega }[2\rho e^{i\theta }\left( e^{-ik}\beta _{-k}^{\dagger
}\beta _{k}-\alpha _{-k}^{\dagger }\alpha _{k}\right)  \notag \\
&&+\left( 1-e^{i\left( 2\theta -k\right) }\right) \left( \beta _{k}\alpha
_{k}+\beta _{-k}^{\dagger }\alpha _{-k}^{\dagger }\right)  \notag \\
&&+\left( e^{i\left( 2\theta -k\right) }+1\right) \left( \beta
_{-k}^{\dagger }\alpha _{k}+\beta _{k}\alpha _{-k}^{\dagger }\right) ]
\end{eqnarray}%
and%
\begin{equation}
\varepsilon _{\rho \sigma }^{k}+\varepsilon _{\rho \overline{\sigma }}^{k}=2%
\mathrm{Re}\varepsilon _{\rho \sigma }^{k}
\end{equation}%
which are independent of $\Delta _{b}-\Delta _{a}$, with $\overline{\sigma }%
=-\sigma $. It indicates that any pair eigenstate in the form%
\begin{equation}
\left\vert \psi _{\mathrm{pair}}\right\rangle =\overline{A}_{\rho \sigma
}^{k}\overline{A}_{\rho \overline{\sigma }}^{k}\left\vert \mathrm{Vac}%
\right\rangle ,
\end{equation}%
is independent of $\Delta _{b}-\Delta _{a}$ and has real energy 
\begin{equation}
E_{\mathrm{pair}}=2\sum_{\{k\}}\mathrm{Re}\varepsilon _{\rho \sigma }^{k}.
\end{equation}%
Here $\left\vert \mathrm{Vac}\right\rangle $\ is the vacuum state of the set
of operators $\{A_{\rho \sigma }^{k}\}$, i.e., $A_{\rho \sigma
}^{k}\left\vert \mathrm{Vac}\right\rangle =0$. On the other hand, the ground
state of $H_{0}+H_{\pi }$, with%
\begin{eqnarray}
H_{0} &=&2J\alpha _{0}^{\dagger }\beta _{0}+2J\beta _{0}^{\dagger }\alpha
_{0}  \notag \\
&&+\left( \Delta _{b}-\Delta _{a}\right) \left( \alpha _{0}^{\dagger }\beta
_{0}^{\dagger }+\alpha _{0}\beta _{0}\right) ,  \notag \\
H_{\pi } &=&\left( \Delta _{b}+\Delta _{a}\right) \left( \alpha _{\pi
}^{\dagger }\beta _{\pi }^{\dagger }+\beta _{\pi }\alpha _{\pi }\right) ,
\end{eqnarray}%
can be easily obtained by diagonalization, yielding%
\begin{equation}
\left( H_{0}+H_{\pi }\right) \left\vert \text{\textrm{g}}\right\rangle
=-(2J+\Delta _{a}+\Delta _{b})\left\vert \text{\textrm{g}}\right\rangle ,
\end{equation}%
with%
\begin{equation}
\left\vert \text{\textrm{g}}\right\rangle =\frac{1}{2}\left( \alpha _{\pi
}^{\dagger }\beta _{\pi }^{\dagger }+1\right) \left( \alpha _{0}^{\dagger
}-\beta _{0}^{\dagger }\right) \left\vert 0\right\rangle .
\end{equation}%
where $\left\vert 0\right\rangle $\ is the vacuum state of the fermion
operators. Obviously, the ground state is enclosed in the set of eigenstates
in the form $\left\vert \psi _{\mathrm{pair}}\right\rangle \left\vert \text{%
\textrm{g}}\right\rangle $.

\end{document}